\documentclass[12pt]{article}
\usepackage[utf8]{inputenc}
\usepackage{graphicx}
\usepackage{natbib}
\usepackage{color}
\usepackage{amsmath}
\usepackage{amsfonts}
\usepackage{hyperref}
\usepackage[algo2e]{algorithm2e} 
\usepackage{algorithm}
\usepackage{verbatim}
\usepackage{caption}
\usepackage{subcaption}
\usepackage[left=1in,right=1in,top=1in,bottom=1in]{geometry}
\newcommand{\bc}{\boldsymbol c}
\newcommand{\bp}{\boldsymbol p}

\title{Quantifying the presence/absence of meso-scale structures in networks}
\author{Eric Yanchenko
\\ Department of Statistics\\ North Carolina State University, Raleigh, NC 27606}

\begin{document}

\maketitle
\thispagestyle{empty}

\begin{abstract}
\noindent
Meso-scale structures are network features where nodes with similar properties are grouped together instead of being treated individually. In this work, we provide formal and mathematical definitions of three such structures: assortative communities, disassortative communities and core-periphery. We then leverage these definitions and a Bayesian framework to quantify the presence/absence of each structure in a network. This allows for probabilistic statements about the network structure as well as uncertainty estimates of the group labels and edge probabilities. The method is applied to real-world networks, yielding provocative results about well-known network data sets.
\end{abstract}

\noindent
{\it Keywords:} Assortativity, Bayesian inference, Community structure, Core-periphery, Disassortativity, MCMC, Model selection

\clearpage

\section{Introduction}
\begin{figure}
    \centering
    \begin{subfigure}{0.33\textwidth}
        \centering
        \includegraphics[width=0.9\textwidth]{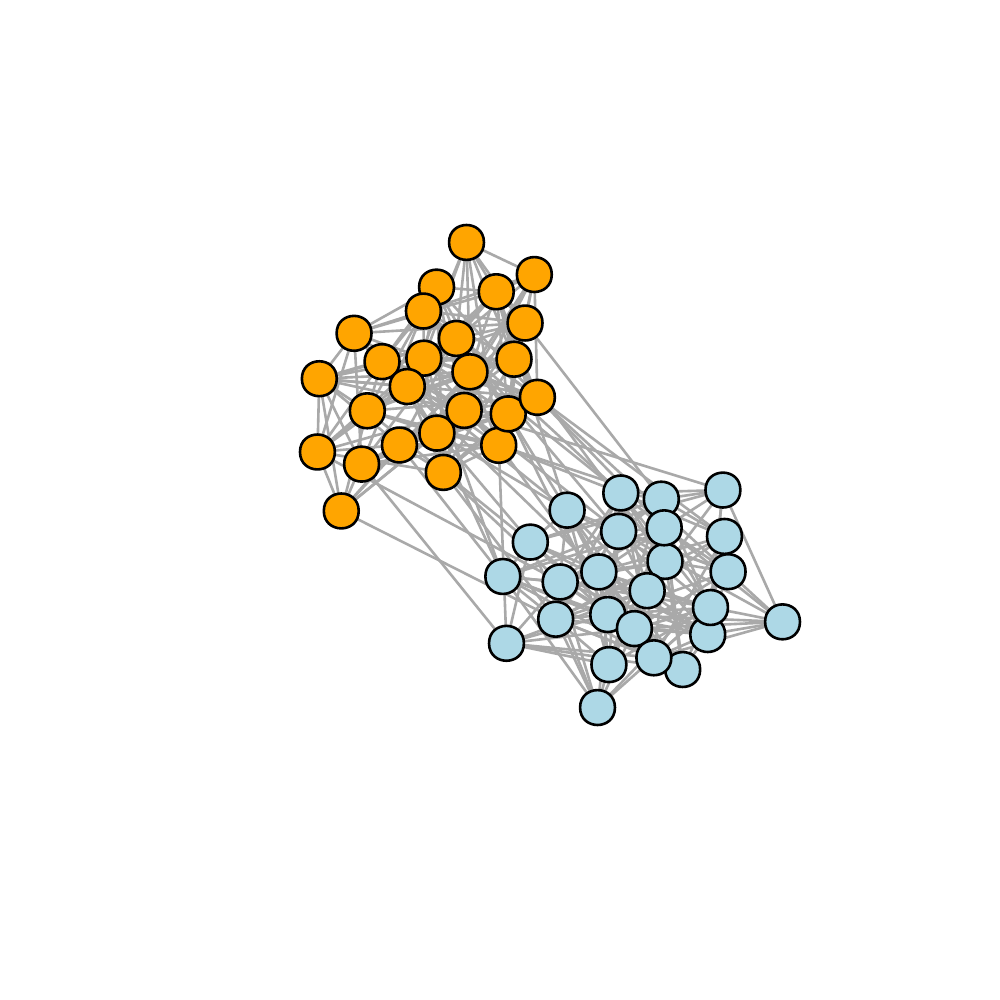} 
        \caption{Assortative communities}
    \end{subfigure}\hfill
        \begin{subfigure}{0.33\textwidth}
        \centering
        \includegraphics[width=0.9\textwidth]{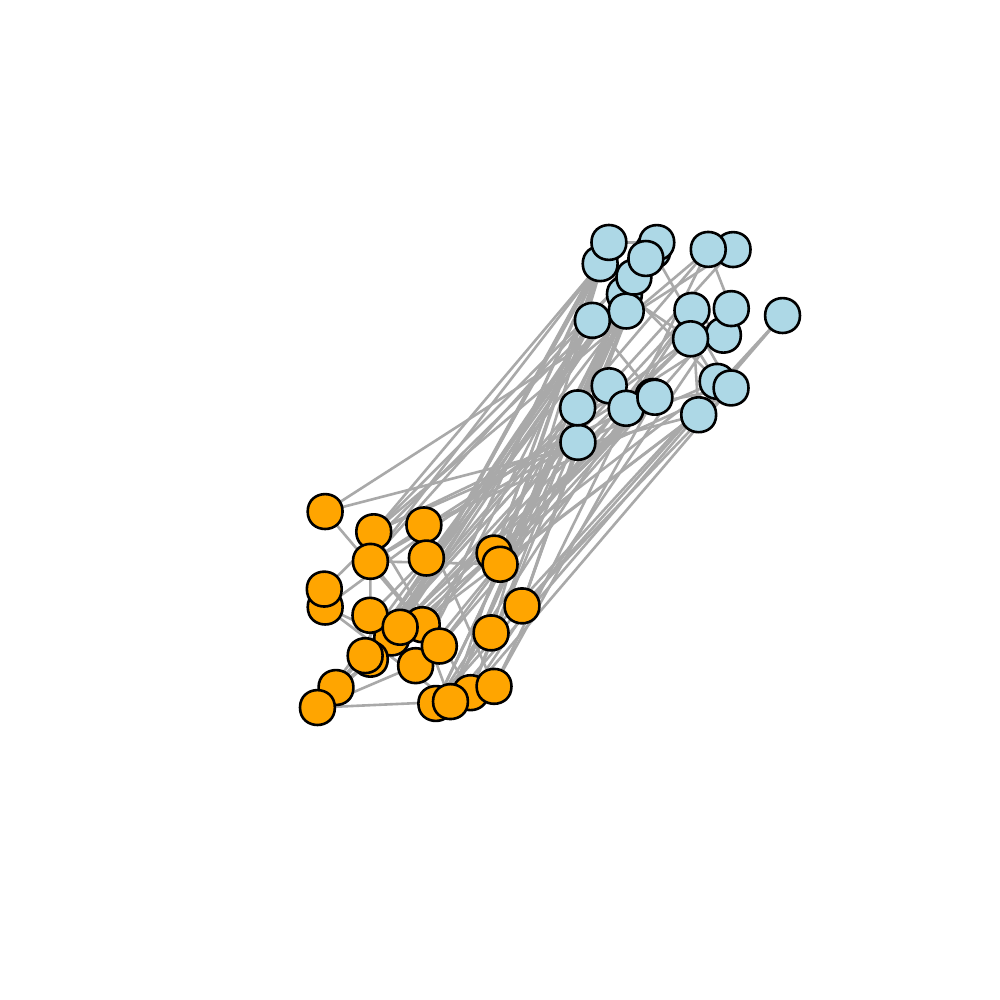} 
        \caption{Disassortative communities}
    \end{subfigure}\hfill
    \begin{subfigure}{0.33\textwidth}
        \centering
        \includegraphics[width=0.9\textwidth]{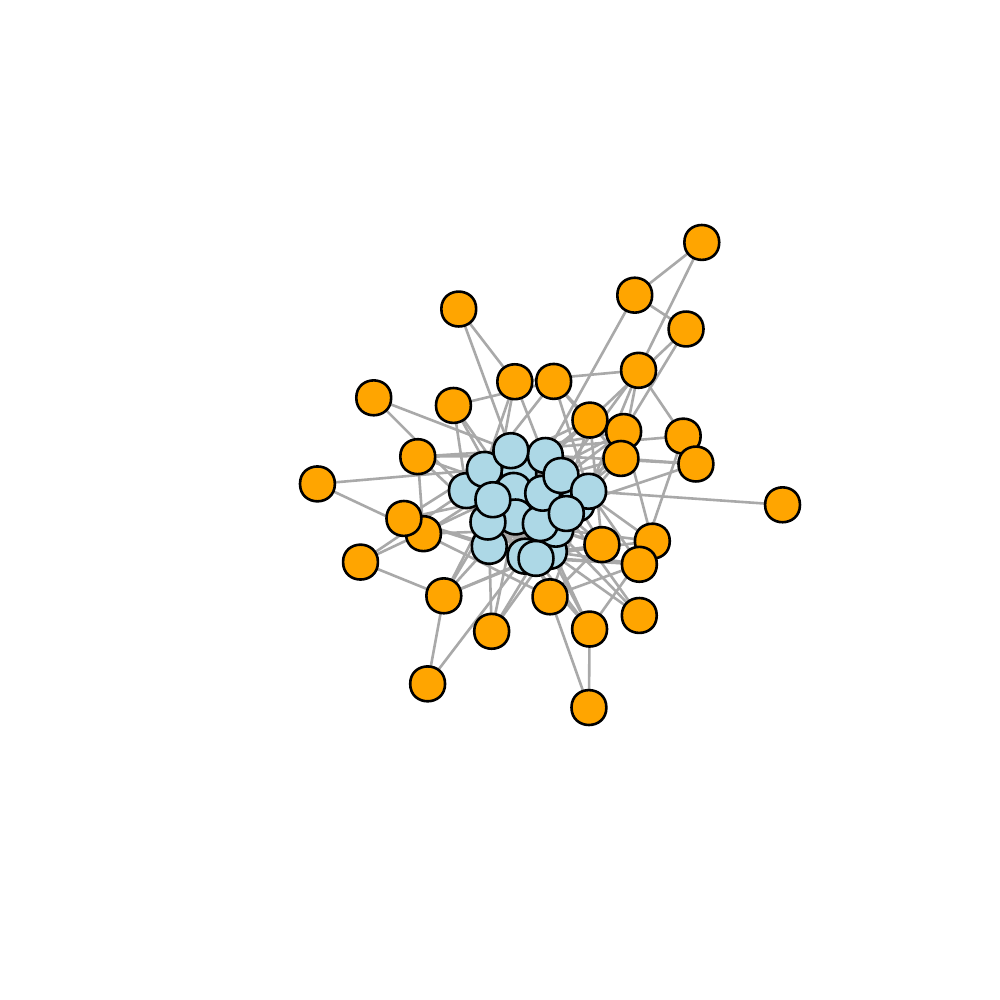} 
        \caption{Core-periphery}
    \end{subfigure}
    \caption{Comparison of networks generated with assortative/disassortative community structure and core-periphery structure.}
    \label{fig:ex}
\end{figure}

One of the most prominent areas of networks research has been on {\it meso-scale structures}. As to opposed to treating individual nodes as the ``building blocks" of a network, meso-scale structures treats groups of nodes as the units of interest. Thus, the network can be viewed from the ``medium" or ``meso" scale. By far the most studied meso-scale feature is {\it community structure} where communities form the groups of nodes. {\it Assortative community structure} occurs when nodes are highly connected within communities and loosely connected between communities \citep{mcpherson2001birds, Newman:2004aa, Newman:2006aa}. {\it Disassortative community structure} is the opposite, where nodes are highly connected between communities and loosely connected within \citep{Fortunato:2010aa, newman2018networks}. Another well-known meso-scale structure is {\it core-periphery structure} which consists of a densely connected core and a loosely connected periphery \citep{BORGATTI2000, Csermely2013, yanchenko2022coreperiphery}. Figure \ref{fig:ex} highlights these similarities and differences between the three structures.

The majority of research on meso-scale structures has studied these properties in {\it isolation}. In other words, most methods assume a particular feature of interest into the model or algorithm. For example, many graph partitioning algorithms maximize the number of within-group edges, implicitly assuming assortative community structure \citep[e.g.,][]{Kernighan1970}. In practice, however, the network structure is unknown before the analysis. Thus, forcing a structure onto the data may lead to invalid claims which can have important, real-world significance. 

To see the potential ramifications of this model misspecification, consider a world trade network where nodes are countries and edges represent trade between the countries and assume that a recession occurs in one country.
If the network has CP structure, then this recession would have a devastating impact on the entire world economy if it is a core node but only a minimal impact if it is a periphery node. If the network has assortative community structure, then only the nodes in this country's community would be hurt while the economies in the other countries would be relatively unaffected. 
Thus, if we (incorrectly) modeled this network with assortative communities when in reality the network exhibited CP structure (or vice-versa), we may draw misleading conclusions on the impact of the recession. 

The goal of this work, then, is to develop a general, data-driven approach to quantify the presence/absence of each meso-scale structure in a given network.
While this particular question seems to have garnered minimal attention, there has been some previous work looking at the relationship between different meso-scale structures. \cite{Yang2014} argue that cores arise from the intersection of many overlapping communities. \cite{tuncc2015unifying} provide a unified formulation that allows for a hybrid of community and CP structure. In particular, the edge probability $p_{ij}$ of a node in group $i$ and $j$ is modeled as
\begin{equation}
        p_{ij}
        =a\delta_{\bc_i,\bc_j}(C_i+C_j-C_iC_j) + b
\end{equation}
where $\delta_{xy}$ is the Kronecker delta taking value $1$ if $x=y$ and 0 otherwise, $\bc_i$ is the group of node $i$ and $C_i$ is some measure of ``coreness" for node $i$. Thus, the $\delta_{\bc_i,\bc_j}$ accounts for the community structure and the $C_i+C_j-C_iC_j$ accounts for the CP structure. \cite{yang2018structural} find Twitter networks composed of multiple communities with CP structure within each community. Lastly, \cite{Kojaku2018}, argue that a third block (e.g., community) is needed for CP structure. 

We propose a Bayesian approach to find and compare meso-scale features for a given network. Our method begins with a principled and mathematical definition of each of the three structures. Moreover, no single structure is enforced in the model which allows the data to drive the results. The Bayesian framework provides clear and probabilistic statements about the likelihood of each structure while also quantifying the uncertainty of group labels and edge probabilities. We also present visualization tools to aid in inference. In Section 2, we propose the model which allows for definitions of the three structures. We also describe the posterior computation and advantages of the proposed method. We apply the methods to synthetic and real-world data in Section 3 and close in Section 4 with limitations and future work. 

\section{Methodology}
\subsection{Model}
We adopt the model of \cite{snijders1997} and \cite{Newmann2015}. Assume that $G=(E,V)$ is a simple, undirected, unweighted network and let $A$ be the associated $n\times n$ adjacency matrix where $A_{ij}=1$ if nodes $i$ and $j$ have an edge and 0 otherwise. Let $A\sim P$ be shorthand for $A_{ij}\sim\mathsf{Bernoulli}(P_{ij})$ for $1\leq i<j\leq n$ and conditioned on $P$, the entries of $A$ are independent. Consider a two-block Stochastic Block Model (SBM) \citep{holland83} with $n\times 1$ labels vector $\bc$ where $c_i=1$ if node $i$ is in group 1 and $c_i=2$ if in group 2 for $i=1,\dots,n$. For CP structure, group 1 is the core and group 2 is the periphery and for assortative/disassortative structure, the two groups are simply two communities. Let $p_{ij}$ be the probability of an edge between nodes in group $i$ and $j$ for $i,j=1,2$ and $\bp=(p_{11},p_{12},p_{22})$. Then the likelihood is
\begin{align}\label{eq:like}
    P(A;\bp,\bc)
    &=\prod_{i<j} p_{c_ic_j}^{A_{ij}}(1-p_{c_ic_j})^{1-A_{ij}}\\
    &=p_{11}^{M_{11}}(1-p_{11})^{m_{11}-M_{11}}p_{12}^{M_{12}}(1-p_{12})^{m_{12}-M_{12}}
    p_{22}^{M_{22}}(1-p_{22})^{m_{22}-M_{22}}.
\end{align}
where $M_{ij}$ is the realized number of edges between block $i$ and $j$ and $m_{ij}$ is the total possible number of edges between block $i$ and $j$. If $n_j=\sum_{i=1}^n \mathbb I(c_i=j)$ is the number of nodes in block $j$, then $m_{ij}=n_i(n_i-1)/2$ if $i=j$ and $m_{ij}=n_in_j$ if $i\neq j$. Notice that both $M_{ij}$ and $m_{ij}$ depend on $A$ and $\bc$ but we suppress this in the notation for convenience. While the two-block SBM can handle the case of no meso-scale features \citep[i.e., Erd\"{o}s-R\'{e}nyi model of][]{erdos1959}, it is a rather strong assumption since it cannot capture degree heterogeneity nor allow for multiple communities. We view these extensions as important areas of future work.

With this formulation, we can precisely and mathematically delineate between the three meso-scale features. We define a network to have assortative community structure if $p_{11}\geq p_{12}$ and $p_{22}\geq p_{12}$; disassortative community structure if $p_{12}\geq p_{11}$ and $p_{12}\geq p_{22}$; and core-periphery structure if $p_{11}\geq p_{12}\geq p_{22}$ where these are based on standard ideas from the literature \citep[e.g.,][]{Fortunato:2010aa, Newmann2015, newman2018networks}. For identifiability, we always let $p_{11}\geq p_{22}$. Notice that the model in (\ref{eq:like}) doesn't enforce or ``hard-code" one particular structure which means it is flexible enough to capture any of them and/or determine which feature is most prominent.

Now, Bayesian methodology performs inference on parameters using Bayes theorem which states
\begin{equation}
    p(\bc, \bp|A)
    \propto p(A|\bc,\bp)\pi(\bc,\bp).
\end{equation}
where $(\bc,\bp)$ are the parameters and $A$ is the data. Thus, we must select prior distributions for both $\bc$ and $\bp$ to compute the posterior distribution. In order not to enforce any meso-scale feature on the network, we chose a prior distribution for $\bp$ such that
\begin{equation}
    \pi(\bp)
    =\pi(p_{11})\pi(p_{12})\pi(p_{22}),
\end{equation}
or, in other words, $p_{11},p_{12}$ and $p_{22}$ are independent {\it a priori}. For conjugacy purposes, we choose a beta prior distribution on each component, i.e., $p_{ij}\sim\mathsf{Beta}(a_{0_{ij}},b_{0_{ij}})$ for $i,j=1,2$ where $X\sim\mathsf{Beta}(a,b)$ has probability density function
\begin{equation}
    f(x;a,b)
    =\frac{\Gamma(a+b)}{\Gamma(a)\Gamma(b)}x^{a-1}(1-x)^{b-1},\ 0\leq x\leq 1
\end{equation} 
where $\Gamma(\cdot)$ is the gamma function. Additionally, our prior distribution for $\bc$ models each term independently such that
\begin{equation}
    \pi(\bc)
    =\pi(c_1)\cdots \pi(c_n)
\end{equation}
where $\pi(c_i=1)=1-\pi(c_i=2)=\pi_i$. Lastly, we assume that $\pi(\bc,\bp)=\pi(\bc)\pi(\bp)$.

\subsection{Posterior computation}
A standard approach to make draws from the posterior distribution is to use a {\it Markov Chain Monte Carlo} (MCMC) sampler. MCMC is a group of sampling methods (e.g., Gibbs and Metropolis-Hastings) used to compute summary statistics of the posterior distribution like the mean and variance \citep{reich2019bayesian}. Under mild assumptions, these estimates converge to the true value of the parameter. With the likelihood and prior distributions specified, we propose the following MCMC routine which uses label swapping to sample $\bc$ and a Gibbs sampler for $\bp$:
\begin{enumerate}
    \item Set initial values of $\bp$ and $\bc$.
    \item For a random ordering of $i=1,\dots,n$, swap the label of $\bc_i$ with probability
    \begin{equation}\label{eq:swap}
        \min\left(1,\frac{P(A;\bc^{(i)},\bp)\pi(c^{(i)}_i)}{P(A;\bc,\bp)\pi(c_i)}\right)
    \end{equation}
    where $c^{(i)}_j=3-\bc_j$ for $j=i$ and $c^{(i)}_j=c_j$ for $j\neq i$.
    \item Sample $\bp|A,\bc$ by drawing $p_{ij}\sim\mathsf{Beta}(M_{ij}+a_{0_{ij}}, m_{ij}-M_{ij}+b_{0_{ij}})$ for\\ $(i,j)\in\{(1,1), (1,2), (2,2)\}$.
    \item Repeat steps (2)-(3) a large number of times.
\end{enumerate}

\subsection{Contributions}\label{sec:inf}
The main contribution of this method is that it simultaneously yields the posterior probability of each meso-scale feature. This allows for statements such as, there is a 70\% chance that the given network has CP structure, 25\% chance it has assortative mixing and 5\% chance of disassortative mixing. The probability that the network has CP structure $P(p_{11}\geq p_{12}\geq p_{22}|A,\bc)$, for example, is found by counting the number of samples with $p_{11}\geq p_{12}\geq p_{22}$ and then dividing by the total number of MCMC samples. The other probabilities are found similarly. Thus, the method gives a simple and interpretable metric to compare the likelihood of each meso-scale features. Most existing methods assume the structure that they are looking for into the model which may lead to false or missed discoveries. The proposed method, however, is flexible and general to allow the data to quantify the features, thanks to the independent priors on $\bp$. This is a key difference from \cite{Gallagher2021}, for example, which selects a prior distribution for $\bp$ that forces a CP relationship into the parameters. While providing probabilities for each meso-scale feature, our approach also gives a sense of significance for the feature by considering the posterior distribution of $\bp$. A sizeable overlap in the distributions of $p_{ij}$ means that this is likely a weak feature whereas a pronounced separation in the distributions implies a more significant structure. Indeed, plotting the posterior distributions provides a novel visual tool to aid in inference.

Additionally, the Bayesian framework gives a measure of uncertainty on the node labels as the relative frequency of $\{c_i=1\}$ is the posterior probability of node $i$ being in group $1$. This gives meaningful insights into the network as a node that has a 55\% probability of being in group 1 is likely different from a node with 95\% probability, for example. Although a frequentist method might assign both methods to group 1, the proposed approach yields this additional information on the node. Moreover, the probability that two (or more) nodes are in the same group can be computed as well as the posterior distribution of each group's size, something that may be of interest in certain applications.

\section{Experiments}\label{sec:data}

\subsection{Synthetic data}
We provide a brief simulation study to demonstrate the performance of the proposed method. For $n=100$ nodes, we generate networks with two blocks containing 40\% and 60\% of the nodes and edge probabilities $p_{11}=0.20$, $p_{22}=0.10$ and $p_{12}\in\{0.05,0.075,\dots,0.25\}$. Thus, $p_{12}=0.10$ is the threshold from assortative communities to core-periphery structure and $p_{12}=0.20$ is the threshold from core-periphery structure to disassortative communities. For each setting, we generate 100 networks and compute the average posterior probability of each meso-scale feature. We collect 1500 MCMC samples with the first 500 discarded as burn-in.

\begin{figure}
    \centering
    \includegraphics[scale=0.85]{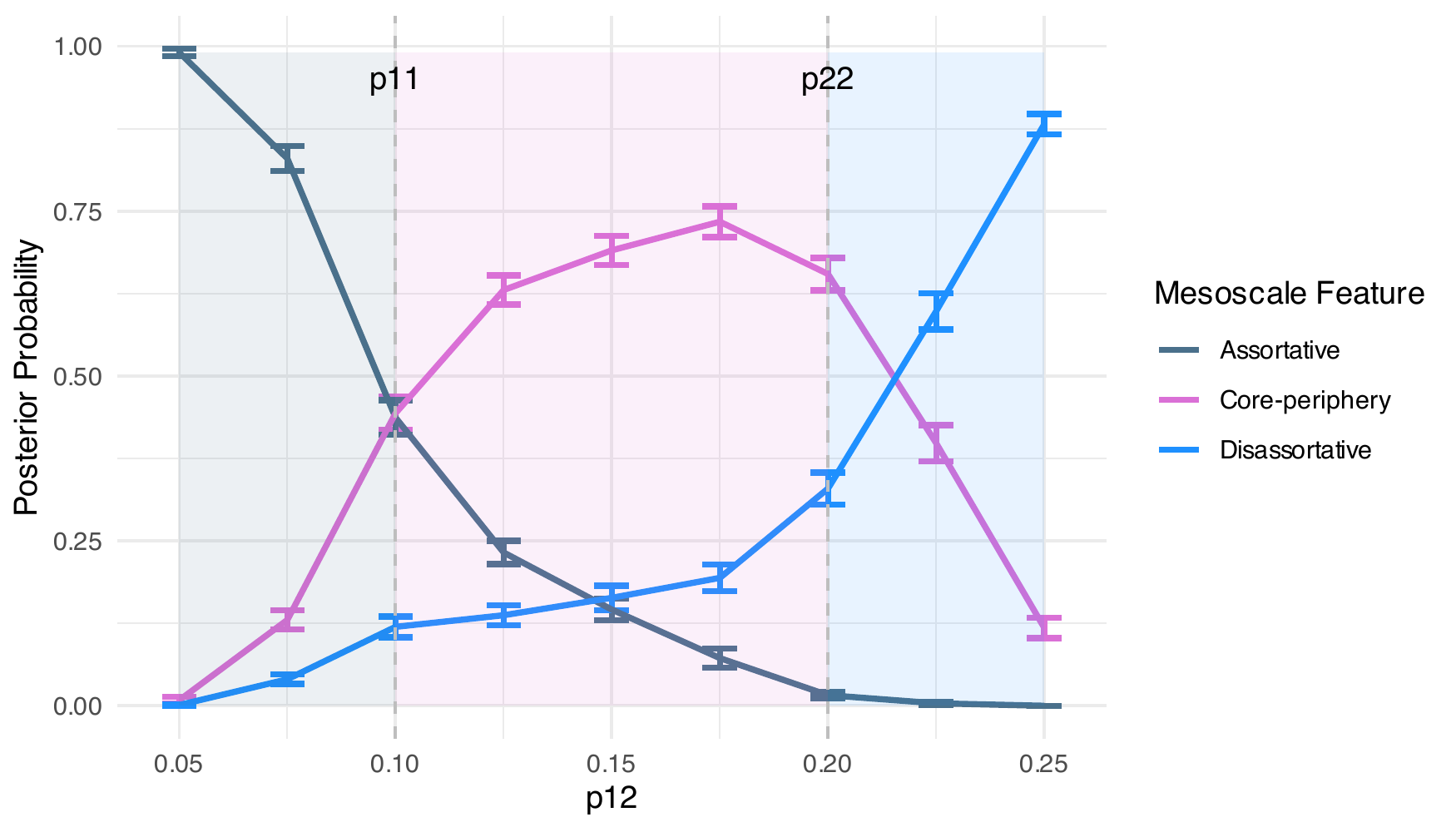}
    \caption{Posterior probability of each meso-scale feature. Vertical lines at $p_{11}=0.10$ and $p_{22}=0.20$. Error bars are the standard error of the means.}
    \label{fig:sim}
\end{figure}

The results are plotted in Figure \ref{fig:sim}. In the assortative community structure region ($0.05\leq p_{12}\leq 0.10$), this feature has the largest average posterior probability. As the networks shift to exhibiting core-periphery structure ($0.10\leq p_{12}\leq 0.20$), this feature now becomes the most probable with the same trend for networks with disassorative communities ($0.20\leq p_{12}\leq 0.25$). Moreover, the assortative structure posterior probability is monotonically decreasing whereas that of disassortative structure is monotonically increasing, both of which make sense. Additionally, the method gives almost equal probability of assorative community and core-periphery structure when $p_{22}=0.10$ since this is the theoretical ``tipping point" between these two features. These results demonstrate that the proposed method can accurately quantify each meso-scale feature and serves as a good confirmation that the method performs as expected.

\subsection{Real-world data}

We now demonstrate the features of the proposed method on two real-world networks. For each example, we apply our method letting $a_{0_{ij}}=b_{0_{ij}}=1$ for $i,j=1,2$ and $\pi_i=0.5$ for $i=1,\dots,n$. Additionally, 15 000 samples were collected with the first 5 000 discarded as burn-in.

\paragraph{Karate club}
First, we consider the famous Karate club dataset from \cite{zachary:1977}. The $n=34$ nodes are different members of a Karate club and the $m=78$ nodes represent some social relationship between two members. A fission in the group led to a split and start of two new clubs so these are usually considered ``ground-truth" communities. 

\begin{figure}
    \centering
    \includegraphics[scale=0.70]{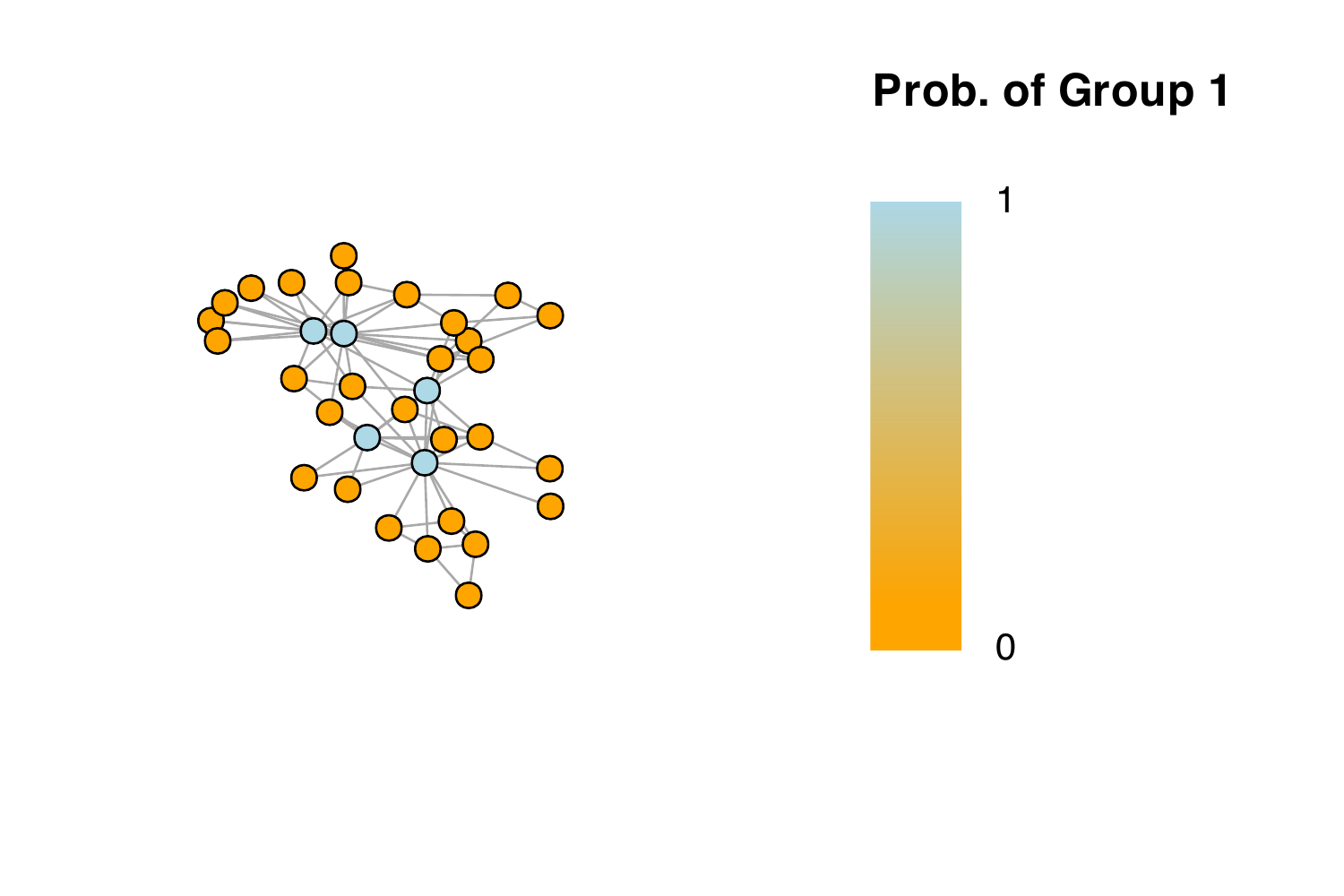}
    \caption{Karate club network with node color corresponding to posterior probability of assignment to group 1.}
    \label{fig:karate_net}
\end{figure}

After fitting the model, we plot the network in Figure \ref{fig:karate_net} where the color of the node corresponds to the probability of being in each group. The majority of nodes (28/34) are more than 99\% likely to be assigned to one group and the remaining six nodes are all greater than 90\%. This means that the we can make this group assignments with high confidence. Interestingly, the method does not return the two communities corresponding to the group split but rather what appears to be CP structure. Indeed, the posterior probability of CP structure is 0.80 whereas the posterior probability of assortative community structure is 0! This implies that the club's structure was marked more by a CP structure than the two groups that split after the fission.

\begin{figure}
    \centering
    \includegraphics[scale=0.70]{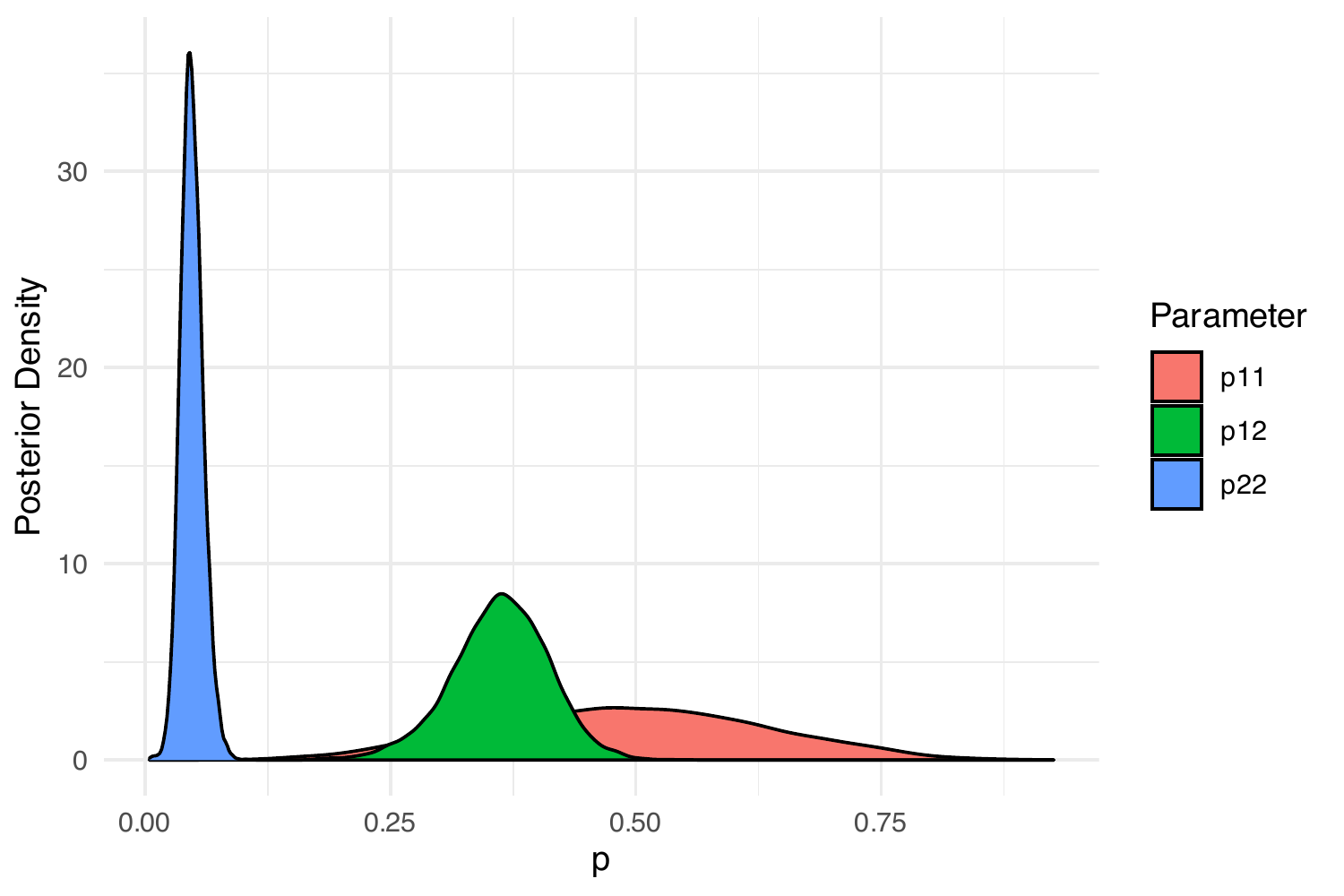}
    \caption{Posterior distribution of $\bp$ for Karate club data.}
    \label{fig:karate_p}
\end{figure}

Next, we look at the posterior distribution of $\bp$ in Figure \ref{fig:karate_p} to determine the significance of this CP structure. There is noticeable overlap between the posterior distribution of $p_{12}$ and $p_{11}$ which means that the CP structure is only moderately strong. The stark separation between $p_{22}$ and $p_{11},p_{12}$, however, indicates no evidence of assortative mixing. This serves as an important example that when no structures are enforced on the data, we may find unexpected results. We note that the lack of community structure in this network was also shown in \cite{yanchenko2021}.

\begin{figure}
    \centering
    \includegraphics[scale=0.70]{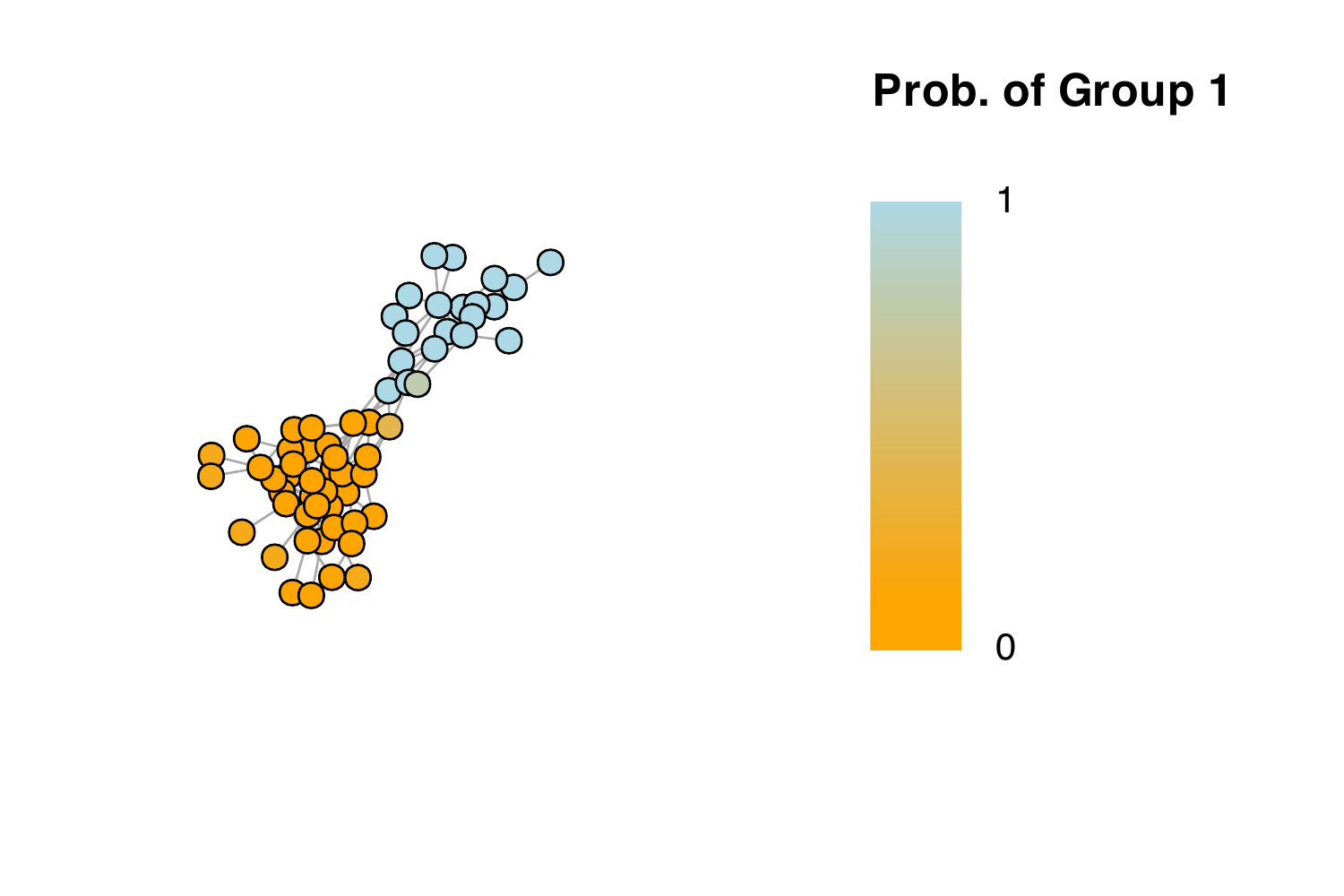}
    \caption{Dolphin network with node color corresponding to posterior probability of assignment to group 1.}
    \label{fig:dolph_net}
\end{figure}

\begin{figure}
    \centering
    \includegraphics[scale=0.70]{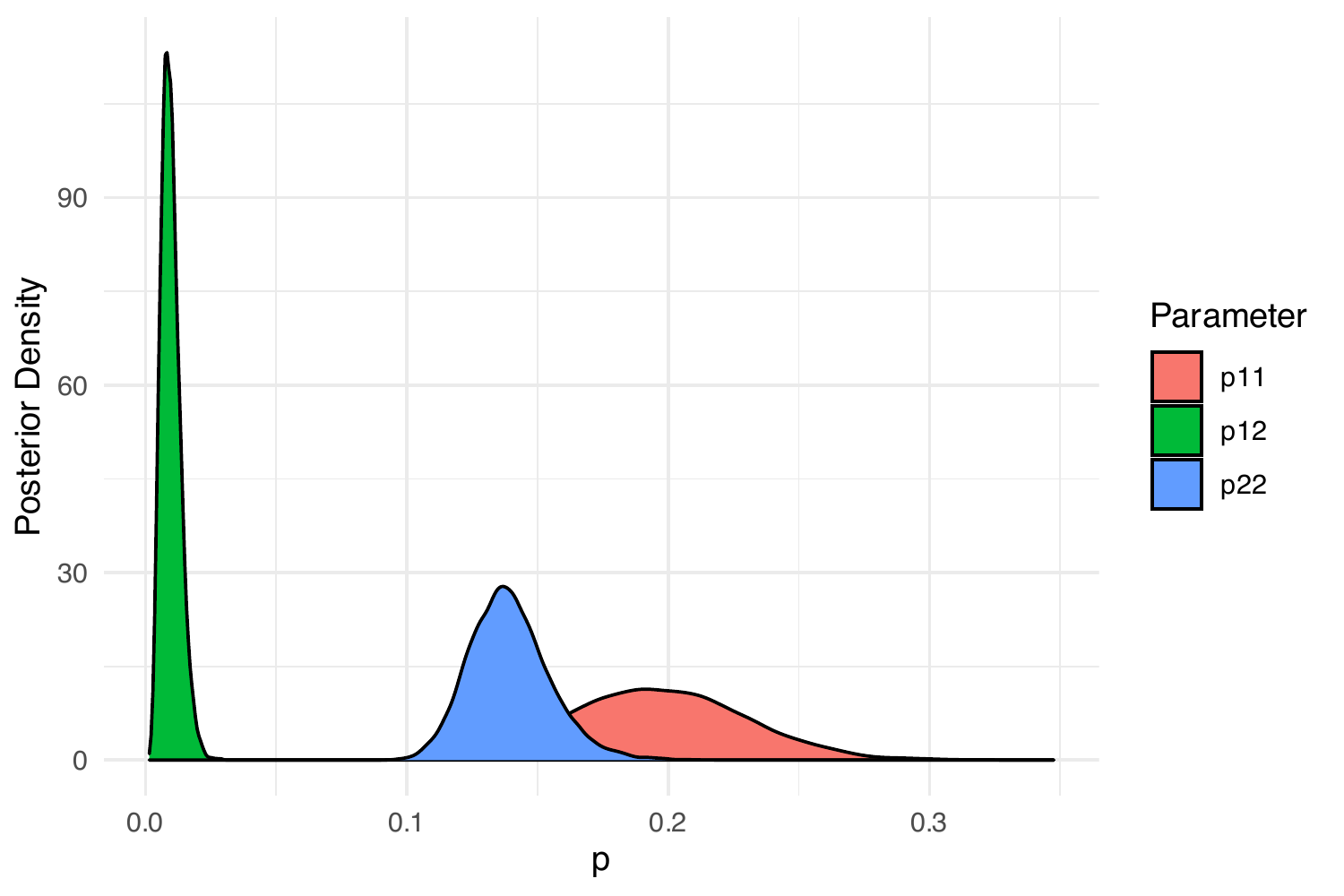}
    \caption{Posterior distribution of $\bp$ for Dolphin data.}
    \label{fig:dolph_p}
\end{figure}

\paragraph{Dolphins}
The second data set comes \cite{lusseau2003bottlenose} who studied the frequent associations between $n=62$ dolphins in Doubtful Sound, New Zealand. There are $m=159$ edges in this network. Many works have hypothesized that this network has assortative community structure \citep[e.g.,][]{Newman:2004aa, Newman:2006aa}.

We plot the network in Figure \ref{fig:dolph_net} where again the color of the node corresponds to the probability of being in each group. The majority of the nodes have clear group assignments, but there are several node which are not neatly assigned to either community (probabilities of 84\%, 84\%, 83\%, 80\%, 68\%). Regardless, the posterior probability of assortative communities is 1. This, coupled with the large separation of the posterior distribution of $p_{11}$ and $p_{22}$ from $p_{12}$ as seen in Figure \ref{fig:dolph_p}, gives very strong evidence for this community structure.

\section{Conclusions}
In this work, we proposed a Bayesian approach to formally quantify the presence/absence of meso-scale features in networks. The strength of the method lies in its ability to make probabilistic statements about the likelihood of each structure as well as yielding uncertainty estimates on community labels. Some limitations of the work are that it requires equal edge probabilities within blocks, something that is likely violated in practice. Thus, an extension similar to degree-corrected stochastic block models \citep{karrer2011stochastic} would be in order. Moreover, the method is currently applicable only for networks with two communities so it could be generalized to multiple communities, as well as weighted and/or directed networks. Finally, another avenue of future work is proving the consistency of the method to recover to the true community labels.
    
\bibliographystyle{apalike}
\bibliography{refs}

\end{document}